\documentclass[a4paper]{article}

\usepackage{times}
\usepackage{fullpage}
\usepackage{amsmath}
\usepackage{amssymb}
\usepackage{graphicx}
\usepackage{clrscode}
\usepackage{float}
\usepackage[square,numbers]{natbib}
\usepackage{authblk}

\graphicspath{{figs/}}

\floatstyle{ruled}
\newfloat{code}{tpbh}{aux}
\floatname{code}{Code}

\title{GPU acceleration of the particle filter: the Metropolis resampler}

\author{Lawrence Murray}
\affil{CSIRO Mathematics, Informatics and Statistics\\
Perth WA, Australia}

\begin{document}

\maketitle

\begin{abstract}
We consider deployment of the particle filter on modern massively parallel
hardware architectures, such as Graphics Processing Units (GPUs), with a focus
on the resampling stage. While standard multinomial and stratified resamplers
require a sum of importance weights computed collectively between threads, a
Metropolis resampler favourably requires only pair-wise ratios between
weights, computed independently by threads, and can be further tuned for
performance by adjusting its number of iterations. While achieving respectable
results for the stratified and multinomial resamplers, we demonstrate that a
Metropolis resampler can be faster where the variance in importance weights is
modest, and so is worth considering in a performance-critical context, such as
particle Markov chain Monte Carlo and real-time applications.
\end{abstract}

\section{Introduction}

This work considers the adaptation of particle
filtering~\cite{Gordon1993,Doucet2001} to modern graphics processing units
(GPUs). Such devices are typical of a trend away from faster clock speeds
toward expanding parallelism to maximise throughput at the processor
level. Capitalising on this wider concurrency for a given application is
rarely trivial, requiring adaptation of current best-practice sequential
algorithms, development of new algorithms, or even revival of old ideas since
bested in a serial context. This work makes such an attempt for the particle
filter, with a focus on resampling strategies.

Our motivation arises from Bayesian inference in large state-space models, in
particular those of marine biogeochemistry~\cite{Evans1985,Jones2010}. The
broad methodology is that of particle Markov chain Monte Carlo
(PMCMC)~\cite{Andrieu2010}, a meld of MCMC over parameters with sequential
Monte Carlo (SMC) over state. For Metropolis-Hastings in parameter space, the
particle filter is a candidate for computing the marginal likelihood (over the
state) of each newly proposed parameter configuration. This requires iteration
of the particle filter, so the use of GPUs to reduce overall runtime is
attractive. Beyond these applications, the material here may be of relevance
to GPU adaptation of bootstrap methods, or in the presence of hard runtime
constraints as in real-time applications.

For some sequence of time points $t = 1,\ldots,T$ and observations at those
times $\mathbf{y}_1,\ldots,\mathbf{y}_T$, the particle filter estimates the
time marginals of a latent state, $p(\mathbf{X}_t|\mathbf{y}_{1:t})$,
proceeding recursively as:
\begin{enumerate}
\item \emph{(Initialisation)} At $t = 0$, draw $P$ samples
  $\mathbf{x}_0^1,\ldots,\mathbf{x}_0^P \sim p(\mathbf{X}_0)$.
\item \emph{(Propagation)} For $t = 1,\ldots,T$, $i = 1,\ldots,P$ and some
  proposal distribution $q(\mathbf{X}_t|\mathbf{X}_{t-1})$, draw
  $\mathbf{x}_t^i \sim q(\mathbf{X}_t|\mathbf{x}_{t-1}^i)$.
\item \emph{(Weighting)} Weight each particle with
  $w_t^i = \frac{p(\mathbf{y}_t|\mathbf{x}_t^i)
    p(\mathbf{x}_t^i|\mathbf{x}_{t-1}^i)
    p(\mathbf{x}_{t-1}^i|\mathbf{y}_{1:t-1})}{q(\mathbf{x}_t^i|\mathbf{x}_{t-1}^i)}$,  so that the weighted sample set $\{\mathbf{x}_t^i,w_t^i\}$ represents the
  filter density $p(\mathbf{X}_t|\mathbf{y}_{1:t})$.
\end{enumerate}

The basic particle filter suffers from \emph{degeneracy} -- the tendency,
after several iterations, to heap all weight on a single particle. The usual
strategy around this is to introduce a further step:

\begin{enumerate}
\item[4.] \emph{(Resampling)} Redraw, with replacement, $P$ samples from the
  weighted sample set $\{\mathbf{x}_t^i,w_t^i\}$, using weights as
  unnormalised probabilities, and assign a new weight of $1/P$ to each.
\end{enumerate}

Initialisation, propagation and weighting are very readily parallelised,
noting that these are independent operations on each particle. Resampling, on
the other hand, may require synchronisation and collective operations across
particles, particularly to sum weights for normalisation. The focus of this
work is on this resampling stage.

Parallel implementation of resampling schemes has been considered before,
largely in the context of distributed memory
parallelism~\cite{Bolic2004,Bolic2005}. Likewise, implementation of the
particle filter on GPUs has been considered~\cite{Lenz2008,Hendeby2007},
although with an emphasis on low-level implementation issues rather than
potential algorithmic adaptations as here.

\section{Resampling algorithms}\label{sec:resampling}

\begin{figure*}[tp]
\centering
\includegraphics[width=\textwidth]{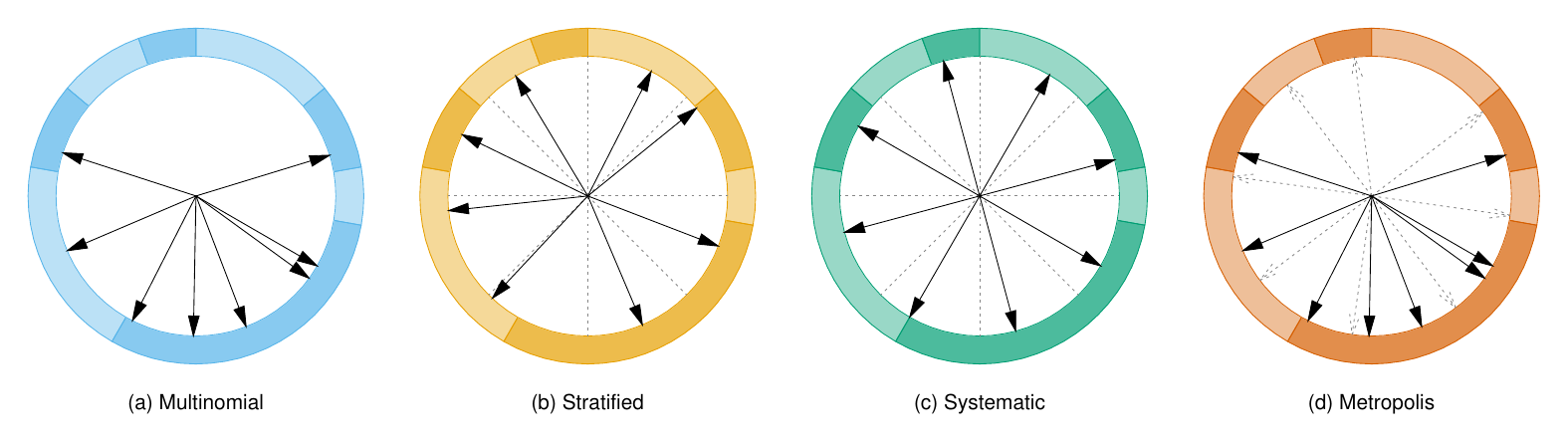}
\caption{Visualisation of popular resampling strategies. Arcs along the
  perimeter of the circles represent particles by weight, arrows indicate
  selected particles and are positioned \textbf{(a)} uniformly randomly in the
  multinomial resampler, \textbf{(b \& c)} by evenly slicing the circle into
  strata and randomly selecting an offset (stratified resampler) or using the
  same offset (systematic resampler) into each stratum, or \textbf{(d)}
  initialising multiple Markov chains and simulating to convergence in the
  Metropolis resampler.}
\label{fig:radial}
\end{figure*}

For each particle $i$ at time $t$, consider the resampling algorithm to be the
means by which its number of \emph{offspring}, $o_i$, for propagation to $t +
1$ is selected, or alternatively the means by which its \emph{ancestor},
$a_i$, from time $t - 1$ is selected. Figure \ref{fig:radial} describes a
number of popular resampling schemes, while Code \ref{code:resamplers}
provides pseudocode.

\begin{code}[t]
{\small
\begin{minipage}{\linewidth}
\input{multinomial.tab}
\end{minipage}
\begin{minipage}[t]{0.5\linewidth}
\input{systematic.tab}
\end{minipage}
\hfill
\begin{minipage}[t]{0.5\linewidth}
\input{metropolis.tab}
\end{minipage}
}
\caption{Pseudocode for the multinomial, systematic and Metropolis resamplers.}\
\label{code:resamplers}
\end{code}

The stratified and systematic~\cite{Kitagawa1996} resamplers output offspring,
while the multinomial and Metropolis resamplers output ancestors. Converting
between the two is straightforward. The multinomial, stratified and systematic
resamplers require a collective prefix-sum of weights. While this can be
performed quite efficiently on GPUs~\cite{Sengupta2008}, it is not ideal,
requiring thread communication and multiple kernel launches. The Metropolis
resampler requires only ratios between weights, so that threads can operate
independently in a single kernel launch; this is more suited to GPU
architectures.

The Metropolis resampler is parameterised by $B$, the number of iterations to
be performed for each particle before settling on its chosen
ancestor. Selecting $B$ is a tradeoff between speed and reliability: while
runtime reduces with fewer steps, the sample will be biased if $B$ is too
small to ensure convergence. Bias may not be such a problem for tracking
applications where performance is judged by outcomes, but will violate
assumptions that lead to unbiased state estimates in a particle MCMC
framework~\cite{Andrieu2010}.

Convergence depends largely on the particle of greatest weight,
$p_{\text{max}}$, being exposed by a sufficient number of proposals that the
probability of it being returned by any one chain approaches $w_{\text{max}} =
\max_i w_i/\sum_j w_j$. Following \citet{Raftery1992}[\S2.1], construct a
binary 0-1 process $Z_n = \delta(U_n = p_{\text{max}})$ over the sequence
$U_n$ generated by a single chain of the Metropolis resampler. It seems
sensible now to require that $P(Z_B = 1\,|\,Z_0)$ be within some $\epsilon$ of
$w_{\text{max}}$. $Z_n$ is a Markov chain with transition matrix given by:
\begin{equation}
T = \left(\begin{array}{cc}
1 - \alpha & \alpha \\
\beta & 1 - \beta
\end{array}\right)\,,
\end{equation}
where $\alpha$ is the probability of transitioning from 1 to 0, and $\beta$
from 0 to 1. As a uniform proposal across all particle indices is used
(\proc{Metropolis-Resampler}, line \ref{line:proposal}), the chance of
selecting $p_{\text{max}}$ is $1/P$, being of greatest weight this will always
be accepted through the Metropolis criterion (line \ref{line:accept}), and so
$\beta = 1/P$. For $\alpha$, we have:
\begin{equation}
\alpha = \sum_{i = 1, i \neq p_{\text{max}}}^{P} \left(\frac{1}{P}
\cdot \frac{w_i/\sum_j w_j}{w_{\text{max}}}\right)
= \frac{1}{Pw_{\text{max}}}(1 - w_{\text{max}})
\,.
\end{equation}
The $l$-step transition matrix is then:
\begin{equation}
T^l = \frac{1}{\alpha + \beta}\left(\begin{array}{cc}
\alpha & \beta \\
\alpha & \beta
\end{array}\right) + \frac{\lambda^l}{\alpha + \beta}\left(\begin{array}{cc}
\alpha & -\alpha \\
-\beta & \beta
\end{array}\right)\,,
\end{equation}
where $\lambda = (1 - \alpha - \beta)$, and we require:
\begin{equation}
\lambda^B \leq \frac{\epsilon(\alpha + \beta)}{\max(\alpha,\beta)}\,,
\end{equation}
satisfied when:
\begin{equation}
B \geq \frac{\log \frac{\epsilon(\alpha + \beta)}{\max(\alpha,\beta)}}{\log
  \lambda}\,.
\end{equation}
In practice one may select $P$ and an upper tolerance for $w_{\text{max}}$
based on expected weight variances given the quality of $q(\cdot)$ proposals,
and compute an appropriate $B$ as above to use throughout the filter.

\section{Experiments}\label{sec:experiments}

Weight sets are simulated to assess the speed and accuracy of each resampling
algorithm. Assume that weights are approximately Dirichlet distributed,
$\mathbf{w} \sim \text{Dir}(\boldsymbol{\alpha})$, with $\alpha_1 = \alpha_2 =
\ldots = \alpha_P = \alpha$. Data sets are simulated for all combinations of
$P = 256,512,\ldots,65536$ and $\alpha = 10,1,.1,.01$.

\begin{figure}[t]
\centering
\includegraphics[width=\textwidth]{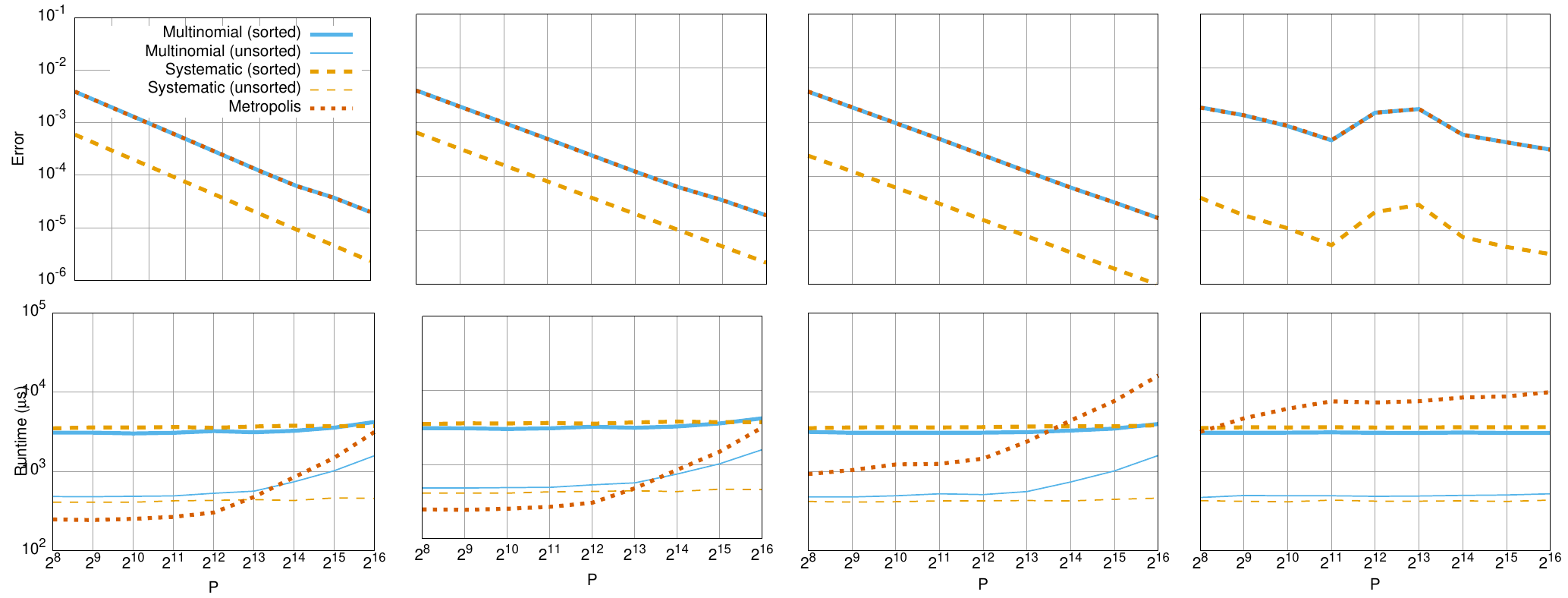}
\caption{Experimental results over 1000 resamplings: \textbf{(top row)} mean
  error, $\sum_{i=1}^{P} (\frac{o_i}{P} - \frac{w_i}{\sum_j w_j})^2$, and
  \textbf{(bottom row)} runtime across various numbers of particles, $P$, with
  Dirichlet $\alpha$ parameter of \textbf{(left to right)} 10, 1, .1 and .01.}
\label{fig:results}
\end{figure}

Experiments are conducted on an NVIDIA Tesla S2050 using double precision,
CUDA 3.1, gcc 4.3 and all compiler optimisations. Implementation of the
Metropolis resampler uses a custom kernel, with random numbers provided by the
Tausworthe~\cite{LEcuyer1996} generator of the Thrust
library~\cite{Thrust}. The multinomial and systematic resamplers use vector
operations of Thrust. To remove the overhead of dynamic memory allocations,
temporaries allocated by Thrust have been replaced with pooled memory, reusing
previously allocated arrays. Figure \ref{fig:results} gives both the accuracy
and runtime of the multinomial, systematic and Metropolis resamplers across
$P$ and $\alpha$.

The Metropolis resampler converges to a multinomial resampling as the number
of steps increases, such that the error in outcomes should match, but can
never beat, that of the multinomial resampler. Figure \ref{fig:results} shows
that the Metropolis resampler has converged in all cases, so certainly the
estimates of $B$ from our analysis appear reliable. The stratified resampler
is known to minimise the variance, and thus error, of resampling outcomes. The
results here confirm this expectation. The multinomial resampler performs
slightly faster than the stratified resampler with sorting enabled, but
slightly slower when not. This is due to the binary search proceeding faster
when weights are sorted. Certainly any thought of pre-sorting to hasten the
binary search seems futile -- the saving is much less than the additional
overhead of the sort on the GPU. Judging by the lack of scaling over $P$,
runtime of the systematic and sorted multinomial resamplers appears dominated
by the overhead of multiple kernel launches.

At $\alpha$ of .1 and .01, the unsorted systematic resampler beats the others
in runtime as well as error. At $\alpha$ of 10 and 1, the Metropolis resampler
is fastest for $P \leq 4096$; in all other cases $B$ must be too large to be
competitive. Note that the modest variance in weights at these $\alpha$ means
that the fairest comparison is against the unsorted approaches, as pre-sorting
is unnecessary for numerical accuracy. At $\alpha = 1$, effective sample size
(ESS) is approximately $.5P$, making this quite a realistic scenario. At
$\alpha = .1$, ESS is about $.1P$, and at $\alpha = .01$ about $.01P$: more
severe cases.

Performance of the Metropolis resampler hinges almost entirely on the rate at
which random numbers can be generated, and this should be the first focus for
further runtime gains. It is possible to reduce $B$, proportionately reducing
runtime, but care should be taken that resampling results are not biased as a
result. Nevertheless, this may be worthwhile under tight performance
constraints, and indeed such configurability might be considered an advantage.

\bibliographystyle{abbrvnat}
{\small

}

\end{document}